\documentclass[aps,prl,twocolumn,showpacs,preprintnumbers,amsmath,amssymb,superscriptaddress,10pt]{revtex4-1}
\usepackage{amsmath,amssymb,amsfonts} 	
\usepackage{graphicx}
\usepackage[english]{babel}
\usepackage[utf8]{inputenc}
\usepackage{subfigure}
\usepackage{xcolor}
\usepackage[normalem]{ulem}
\usepackage{SIunits}
\usepackage[colorlinks=true,linkcolor=blue,citecolor=blue,urlcolor=black]{hyperref}
\makeatletter
\def\bbl@set@language#1{%
  \edef\languagename{%
    \ifnum\escapechar=\expandafter`\string#1\@empty
    \else\string#1\@empty\fi}%
  \@ifundefined{babel@language@alias@\languagename}{}{%
    \edef\languagename{\@nameuse{babel@language@alias@\languagename}}%
  }%
  \select@language{\languagename}%
  \expandafter\ifx\csname date\languagename\endcsname\relax\else
    \if@filesw
      \protected@write\@auxout{}{\string\select@language{\languagename}}%
      \bbl@for\bbl@tempa\BabelContentsFiles{%
        \addtocontents{\bbl@tempa}{\xstring\select@language{\languagename}}}%
      \bbl@usehooks{write}{}%
    \fi
  \fi}
\newcommand{\DeclareLanguageAlias}[2]{%
  \global\@namedef{babel@language@alias@#1}{#2}%
}
\makeatother

\DeclareLanguageAlias{en}{english}

\renewcommand{\[}{\begin{equation}}
\renewcommand{\]}{\end{equation}}
\def\bea{\begin{eqnarray}}
\def\eea{\end{eqnarray}}

\def\runtime{(\the\time)\qquad\the\month/\the\day/\the\year}
\def\today
 {\count10=\year\advance\count10 by -2000 \number\day--\ifcase
  \month \or Jan\or Feb\or Mar\or Apr\or May\or Jun\or
             Jul\or Aug\or Sep\or Oct\or Nov\or Dec\fi--\number\count10}

\def\hour{\count10=\time\count11=\count10
\divide\count10 by 60 \count12=\count10
\multiply\count12 by 60 \advance\count11 by -\count12\count12=0
\number\count10 :\ifnum\count11 < 10 \number\count12\fi\number\count11}

\newcommand\THEOSMARVEL{Theory and Simulation of Materials (THEOS) and National Centre for Computational Design and Discovery of Novel Materials (MARVEL), {\'E}cole Polytechnique F{\'e}d{\'e}rale de Lausanne, 1015, Switzerland}
\begin{document}

\title{Prediction of a large-gap and switchable Kane-Mele quantum spin Hall insulator}

\author{Antimo Marrazzo}
\email{antimo.marrazzo@epfl.ch}
\affiliation{\THEOSMARVEL}
\author{Marco Gibertini}
\affiliation{\THEOSMARVEL}
\author{Davide Campi}
\affiliation{\THEOSMARVEL}
\author{Nicolas Mounet}
\affiliation{\THEOSMARVEL}
\author{Nicola Marzari}
\email{nicola.marzari@epfl.ch}\affiliation{\THEOSMARVEL}

\date{\today}

\begin{abstract}
Fundamental research and technological applications of topological insulators are hindered by the rarity of materials exhibiting a robust topologically non-trivial phase, especially in two dimensions. Here, by means of extensive first-principles calculations, we propose a novel quantum spin Hall insulator with a sizeable band gap of $\sim$0.5 eV that is a monolayer of Jacutingaite, a naturally occurring layered mineral first discovered in 2008 in Brazil and recently synthesised.  This system realises the paradigmatic Kane-Mele model for quantum spin Hall insulators in a potentially exfoliable two-dimensional monolayer,  with helical edge states that are robust and that can be manipulated exploiting a unique strong interplay between spin-orbit coupling, crystal-symmetry breaking and dielectric response. 
\end{abstract}
\maketitle\bigskip \bigskip
The last decade has been marked by a significant effort in the study of topological order in real materials. More than 15 years after the seminal work by Haldane \cite{haldane_88} introducing a model for the Chern insulator (a.k.a. quantum anomalous Hall insulator or QAHI), Kane and Mele \cite{kane_quantum_2005,kane_z2_05} realised that by doubling Haldane's model and introducing spins, they could obtain a quantum spin Hall insulator (QSHI), i.e. a time-reversal invariant insulator characterised by $\mathbb{Z}_2$ topological order and helical edge states \cite{bernevig_book_2013}. Soon, it was recognised that the QSHI is actually a novel state of matter not necessarily bound to the Kane-Mele (KM) model, and the first experimental realisation of a QSHI came in the form of a HgTe/CdTe quantum well \cite{molenkamp_hgte_07}, following a theoretical prediction by Bernevig, Hughes and  Zhang \cite{bernevig_science_06}.  At variance with QAHIs, in QSHIs the non-trivial topological order is protected by time-reversal symmetry (TRS): an even number of Kramers' pair states appears at the edge, potentially hosting dissipation-less electron transport due to the absence of elastic scattering. These counter-propagating edge modes of opposite spin (helical) give rise to topologically protected one-dimensional wires, with the only elastic scattering channel being back-scattering between Kramers pairs, a process totally forbidden by time-reversal symmetry. Thus, helical edge states are very robust against interactions and non-magnetic disorder, making QSHIs a very promising platform to realise novel low-power electronic and spintronic devices. Despite their massive fundamental interest  and their prospective technological applications, experimentally synthesised  QSHIs that persist up to room temperature are still very scarce \cite{zhang_topomatternatmat_17,reis_bisic_17}.

In this work, we first predict by accurate first-principles simulations a novel, optimal QSHI monolayer with a record-high band gap that realises the KM model and that can be extracted from a naturally occurring crystal. Then, we unravel the competing roles of spin-orbit coupling and crystal-symmetry breaking on structural stability, and explore their interplay to show how the topological phase can be switched using moderate, realistic electric fields.
\begin{figure*}[hbtp]
\centering
\includegraphics[width=0.49\textwidth]{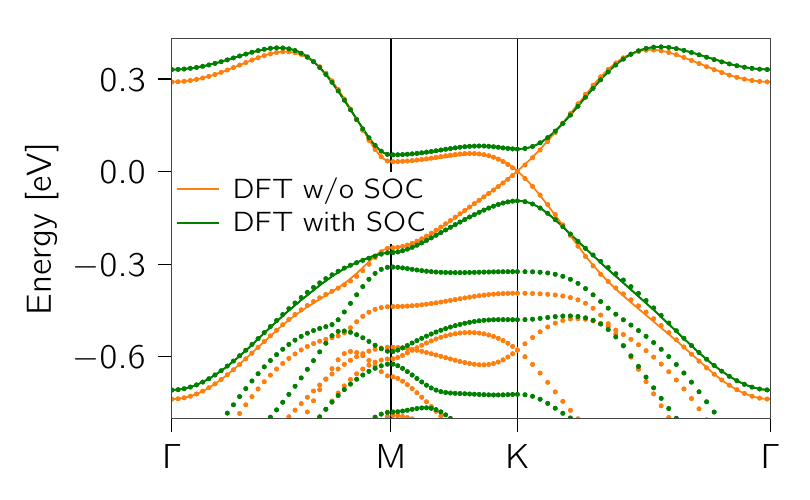}
\includegraphics[width=0.38\textwidth]{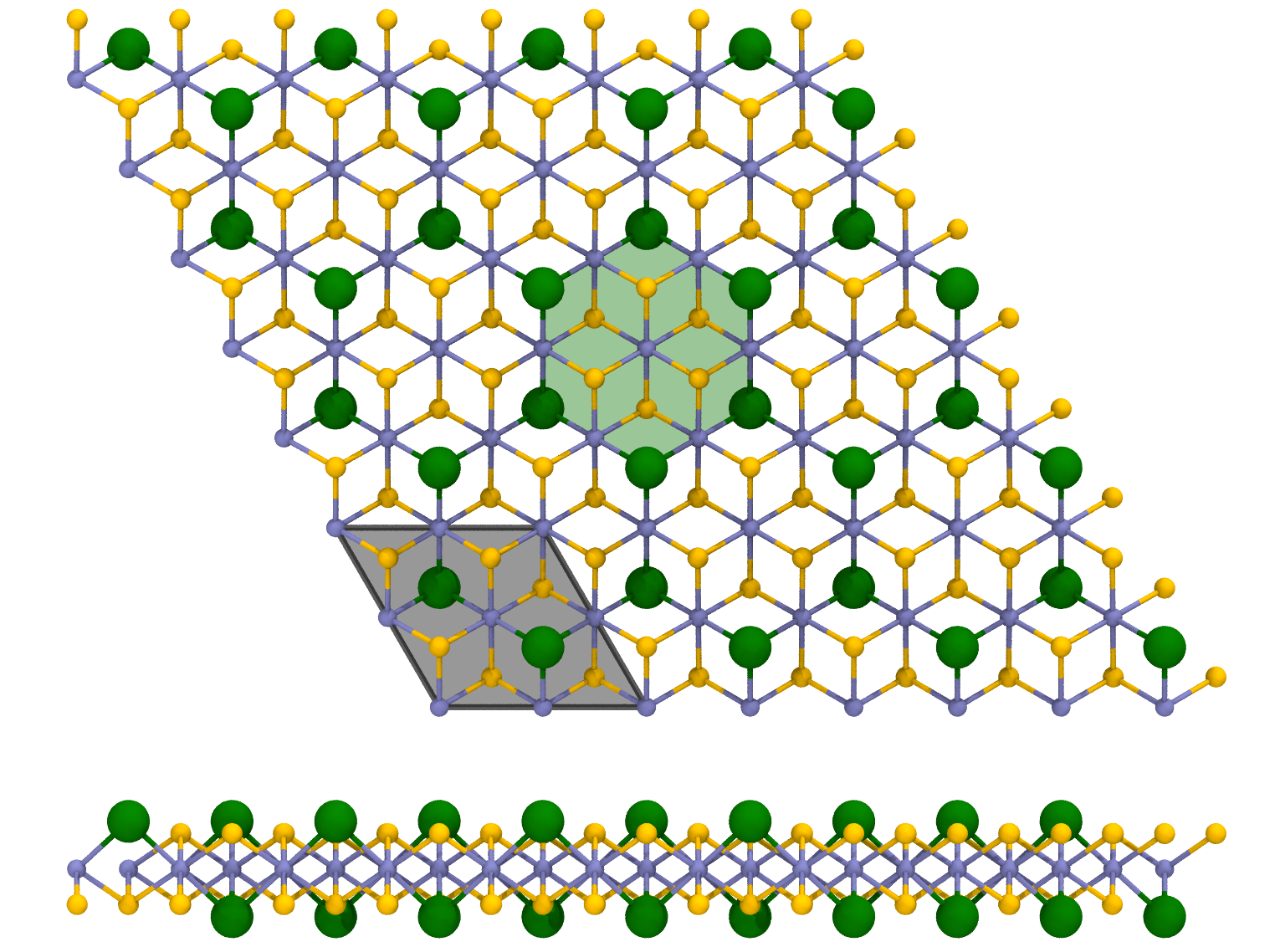}
\includegraphics[width=0.6\textwidth]{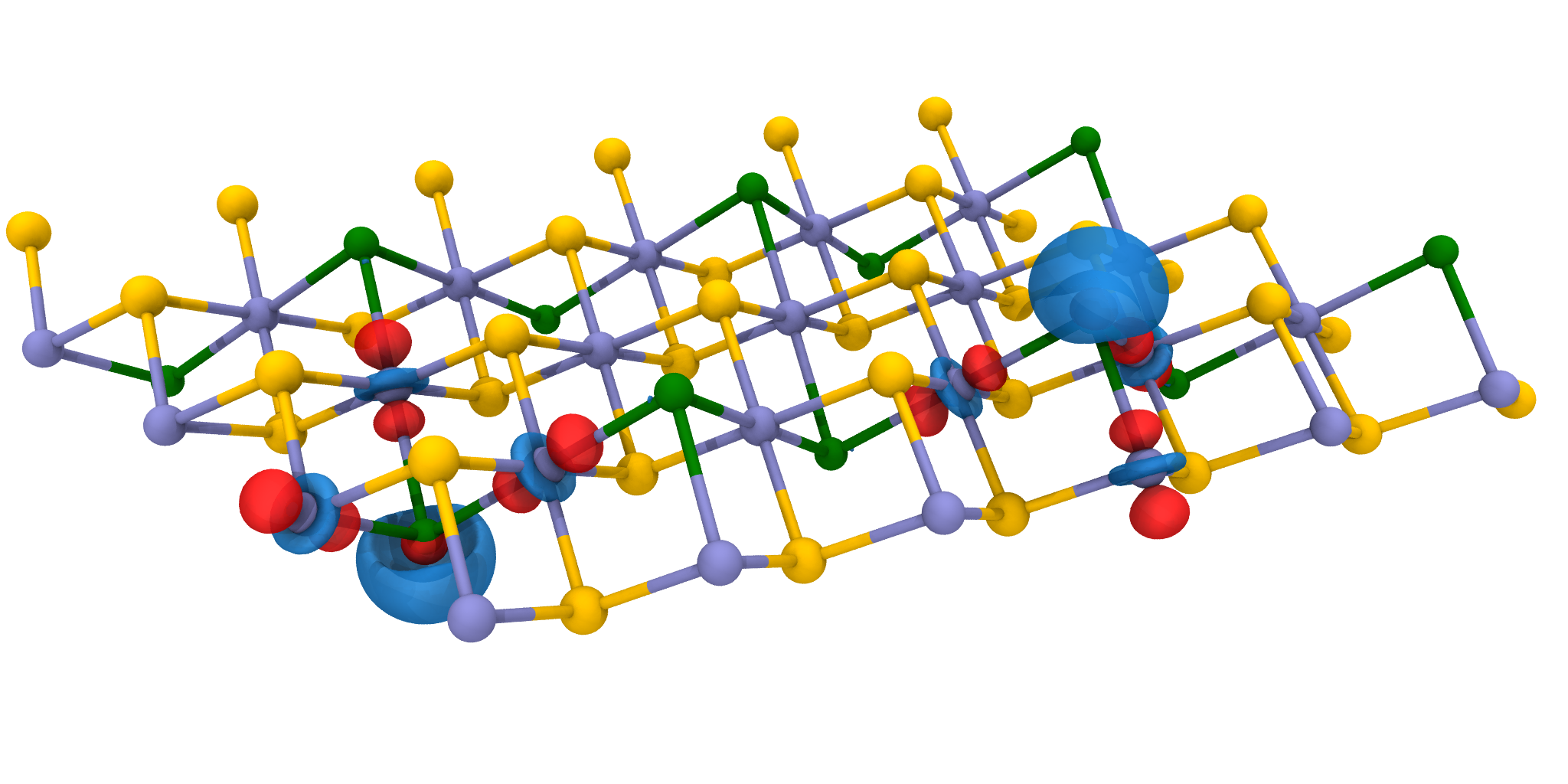}
\caption{\label{fig-model}Top left panel: DFT electronic band structure for monolayer Jacutingaite; the energy zero is the Fermi level of the DFT w/o SOC calculation. Full circles denote direct DFT calculations while solid lines represent Wannier-interpolated states from a minimal two-band model. Green (dark grey) circles and lines represent calculations performed with spin-orbit coupling (SOC), orange (light grey) circles and lines calculations performed without SOC. Top right panel: top and lateral views of monolayer Jacutingaite (Pt$_2$HgSe$_3$); the primitive cell is marked by the black parallelogram while the Wigner-Seitz cell is denoted by the green hexagon. Bottom panel: isosurface plots for the two maximally localised Wannier functions realising the two-band model. The two MLWFs map on to each other under inversion and have the character of Hg $s$-orbitals hybridised with the three nearest-neighbour Pt $d$-orbitals; their centres compose a buckled honeycomb lattice.  Red and blue isosurfaces correspond to opposite signs of the MLWFs. The low-energy physics is fully captured using this two-band model, analogous to the Kane-Mele model for graphene. }
\end{figure*}

Jacutingaite (Pt$_2$HgSe$_3$) is a new species of platinum-group minerals first discovered in 2008 \cite{cabral_first_obs_08}; in 2012, synthetic Jacutingaite was also obtained \cite{jacutingaite_exp_12} and its crystal structure identified with powder X-ray diffraction. The Jacutingaite crystal structure has spacegroup $P\bar{3}m1$ (164), with a trigonal unit cell composed of 12 atoms. The crystal is layered with AA stacking and has a reported \cite{jacutingaite_exp_12} experimental interlayer distance of 5.3 $\angstrom$.  The layered character of Jacutingaite is supported by the experimental reports of ``very good \{001\} cleavage'' for the mineral  \cite{jacutingaite_exp_12}, and a laminated morphology for the synthetic crystals. 
To confirm this, we compute \cite{mounet_nanotech_2017} with non-local van der Waals density-functional theory (vdW-DFT, see Methods in the Supplemental Material \cite{SM_BE}) the geometry and binding energy $E_{\rm b}$ of Jacutingaite finding an interlayer distance of 5.3 $\angstrom$ in exact agreement with experiments, and a binding energy for the monolayer of 60 meV$\cdot$\AA$^{-2}$ \cite{bjorkman_be_12,be_def}. This latter is roughly twice the binding energy obtained \cite{mounet_nanotech_2017} for the recently synthesised CrGeTe$_3$ or for phosphorene, and less than three times the binding energy of graphene or hexagonal boron nitride monolayers, suggesting that monolayer Jacutingaite could be obtained through common exfoliation techniques such as adhesive tape, intercalation or sonication in addition to synthetic growth. The crystal structure of monolayer Jacutingaite is shown in Fig. \ref{fig-model}.
\begin{figure*}[ht]
\centering
\includegraphics{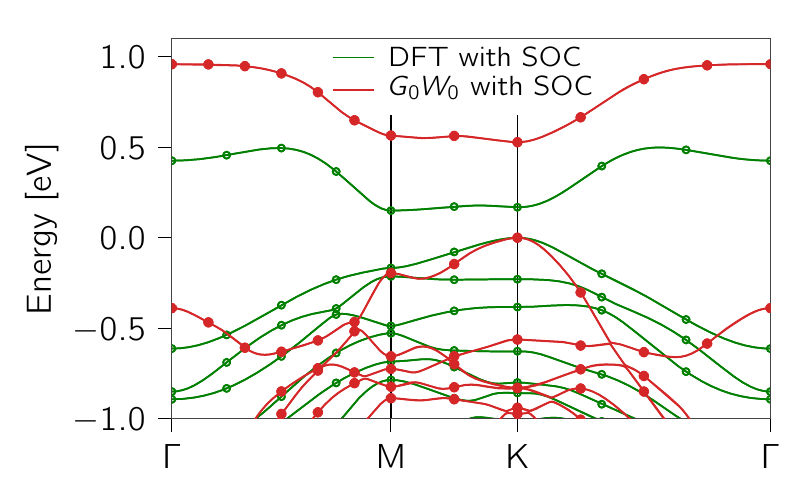}
\includegraphics{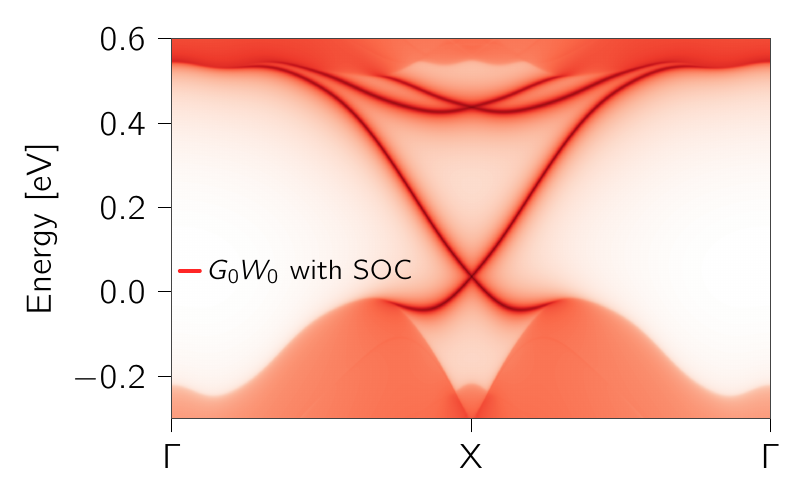}
\caption{\label{fig-gw}Left panel: DFT and $G_0W_0$ band structures obtained including SOC; lines are Wannier-interpolated bands while full ($G_0W_0$) and empty (DFT) circles denote direct calculations. Right panel: $G_0W_0$ edge spectral density displaying a pair of topologically protected helical edge states crossing the bulk gap. }
\end{figure*}
The low-energy physics around the Fermi level can be well described by a two-band model that mirrors the KM model for graphene \cite{kane_quantum_2005}.
 To show this, first we construct an ab initio $2\times2$ Hamiltonian without spin-orbit coupling (SOC) in a basis of maximally-localised Wannier functions (MLWFs) \cite{wannier_review_12}. Fig.~\ref{fig-model} highlights how such a simple model interpolates very well the highest occupied and lowest unoccupied bands as obtained directly from DFT calculations (orange solid lines and circles respectively). The corresponding MLWFs are shown in the bottom-right panel of Fig.~\ref{fig-model}, displaying the character of  Hg $s$-orbitals hybridised with three nearest-neighbour Pt $d$-orbitals. Notably, the MLWFs are centred on the Hg atoms and form a buckled honeycomb lattice, mirroring the structure of germanene or arsenene. Then, we introduce SOC first at the DFT level, and construct again a MLWF Hamiltonian (Fig. \ref{fig-model} top left, green line). In graphene, the only spin-orbit term that respects all symmetries and can open a gap is the KM-type SOC  \cite{kane_quantum_2005}, proportional to $\sigma_z \tau_z s_z$ (these being the Pauli matrices for the sublattice, K/K$^\prime$ valley and spin degrees of freedom, respectively). This strong SOC due to the presence of Hg and Pt gaps the Dirac point and makes the system an insulator with an indirect band gap of 0.15 eV at the DFT level (see later for many-body perturbation theory calculations). In a buckled honeycomb lattice in-plane mirror symmetry is broken, allowing an additional second-nearest neighbour SOC term \cite{liu_buckledSOC_11} that does not affect the band gap at K. In fact, a KM model constructed from first-principles is able to capture the main features of the low-energy bands structure (see Supplemental Material \cite{SM_KM}) and the opening of a gap due to the KM-type SOC. That the system is a QSHI is revealed by calculations of the $\mathbb{Z}_2$ invariant performed by tracking hermaphrodite Wannier charge centres \cite{soluyanov_computing_2011,gresch_z2pack:_2017,sgiarovello_electron_2001} (see Supplemental Fig. 6).

To further assess the robustness of the QSHI phase, we perform many-body $G_0W_0$ calculations with spin-orbit coupling (see Methods). Fig. \ref{fig-gw} shows a comparison between the band structures obtained at the $G_0W_0$ and DFT level. Notably, $G_0W_0$ predicts an expected increase of the band gap to 0.53 eV, with a direct gap at the K point. Such band gap is an order of magnitude higher than the recently synthesised WTe$_2$ \cite{wte2_transp_17,wte2_sts_17,cava_wte2_18}, which is also a monolayer QSHI, albeit driven by orbital band inversion and thus more sensitive to environmental effects.
In Fig. \ref{fig-gw} we show the edge spectral density for a semi-infinite monolayer computed using the iterative Green's function method \cite{lopez_sancho_highly_1985, wu_wanniertools:_2017,lee_nanotubes_05,calzolari_nanotransp_04} on a $G_0W_0$ Wannier Hamiltonian. A pair of edge states crosses at the Fermi level with fairly linear dispersions over all the bulk band gap, an hallmark of $\mathbb{Z}_2$ topological order. 
Such a huge bulk gap in the spectral density should facilitate a clear experimental detection of the QSHI phase, either by scanneling tunneling spectroscopy or transport experiments.
The Fermi velocity estimated from the edge spectral density is rather high ($v_F \approx 3.6 \times 10^{5} m\cdot s^{-1}$), although the precise experimental value depends on possible edge reconstructions. Note that the bulk-boundary correspondence guarantees the presence of helical edge states, independently of the details of the termination as long as TRS is preserved. The large band gap also implies a fast decay of the helical edge states into the bulk, with a transverse localization length approximately equal to \cite{bernevig_book_2013}  $L=\frac{\hbar v_F}{E_{gap}} \approx 5  \angstrom$. Using the $G_0W_0$ bulk Wannier Hamiltonian, we construct nanoribbons of different sizes and confirm that a 6-cells wide ($\sim$4 nm) nanoribbon is sufficient to display gapless helical edge states (see Supplemental Fig. 3).  Hence the two pairs of edge states of a narrow Jacutingaite nanoribbon interact very weakly with each other, suggesting the possibility of realising dissipation-less nanowires in integrated circuits.

\begin{figure*}[!htbp]
\centering
\includegraphics{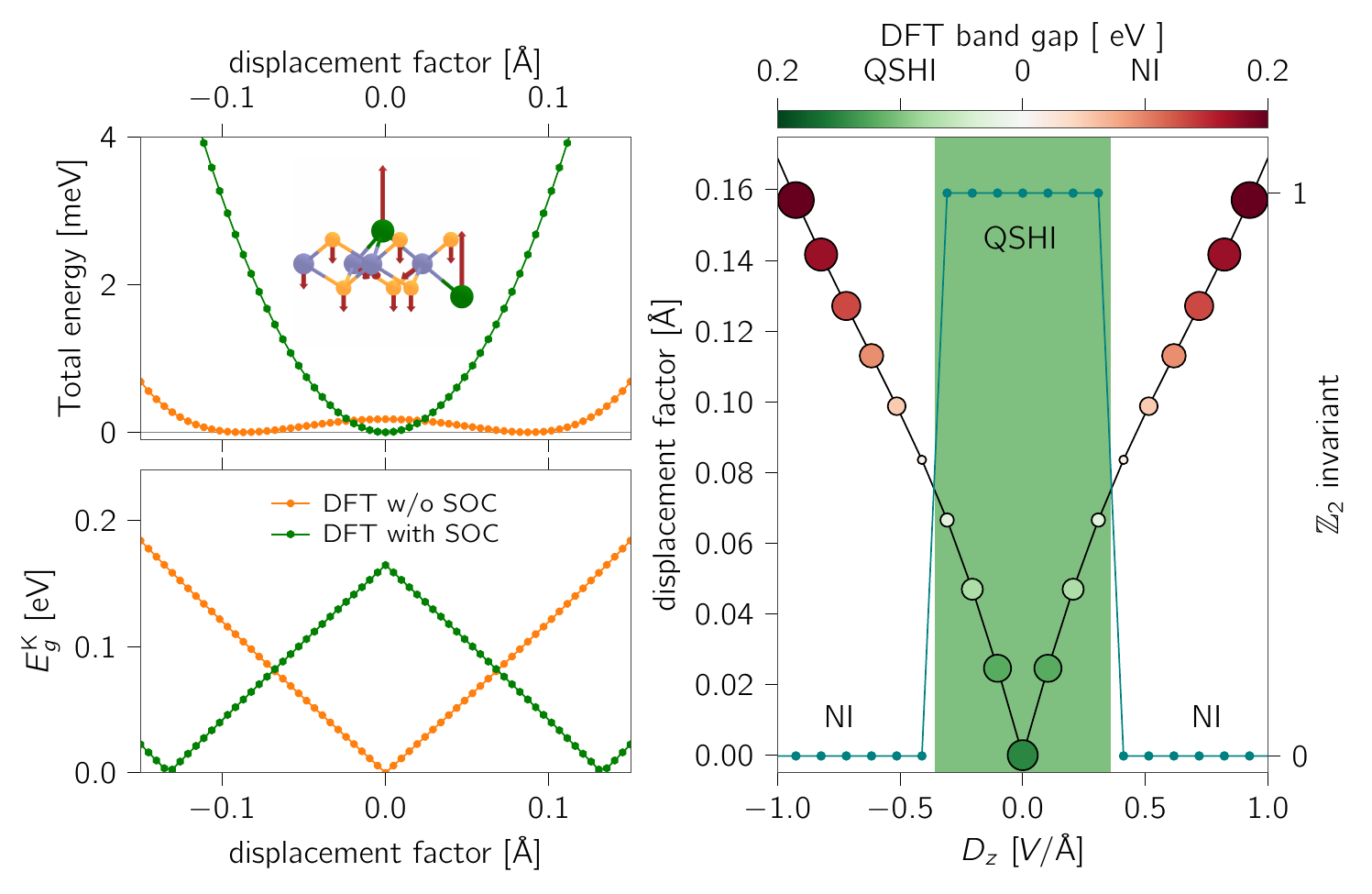}
\caption{\label{fig-stability}Top left panel: DFT total energy vs displacement along a $\Gamma$-unstable displacement pattern, i.e the imaginary-frequency phonon obtained from DFPT calculations performed without SOC (the inset shows the pattern of this phonon calculated in the centro-symmetric configuration, see text). Without including SOC, the system would distort into a polar phase as shown by a double-well potential-energy curve; on the contrary SOC stabilises the centro-symmetric phase and restores a parabolic behaviour. Bottom left panel: direct band gap at K as function of the displacement pattern. In the non-SOC case, inversion symmetry-breaking opens a gap, while including SOC yields gap closure. Right panel: topological phase diagram under an out-of-plane electric displacement field $D_z$. The $\mathbb{Z}_2$ topological invariant is marked in teal; the magnitude of the band gap is represented by the disks' color and size. The black line stands for the projection of the field-induced ionic displacement over the 164-to-156 spacegroup reduction manifold (see text). The ionic displacement due the field acts similarly to the imaginary-frequency phonon, causing an ionic response that reduces the critical field where the topological phase transition occurs.}
\end{figure*}

Now we discuss the mechanical stability of the monolayer. We compute phonon dispersions using 2D DFPT \cite{baroni_dfpt_review_01,sohier_2dcode_17} including SOC and the correct 2D LO-TO asymptotics \cite{sohier_loto_17}  (see Supplemental Material \cite{SM_BE}); the stability of the monolayer is confirmed by the absence of imaginary frequencies. Interestingly, the zero-temperature centro-symmetric phase is promoted by the presence of KM-type SOC. In fact, DFPT calculations without SOC return an instability at $\Gamma$ that would break inversion symmetry and lower the spacegroup from $P\bar{3}m1$ (164) to $P3m1$ (156).  We further investigate this by computing several DFT total energies obtained by displacing the atoms according to the pattern of the unstable phonon. Fig. \ref{fig-stability} illustrates this mechanism: without SOC the system would prefer to distort into one of two equivalent polar phases, as shown by a double-well potential-energy curve, although the energy barrier between the two polar phases would be small compared with room-temperature, and the system would rather stay in a ``quantum paraelectric'' phase  \cite{muller_quantumparaelectric_79,vand_quantumparaelectric_96}. On the contrary, SOC stabilises the centro-symmetric phase and restores a parabolic behaviour for the total energy. This phenomenon can be understood by studying how SOC affects the behaviour of the band gap under the inversion-symmetry breaking distortion (see the bottom-left panel of Fig. \ref{fig-stability}).  Without SOC, the centro-symmetric phase is a Dirac semimetal and the distortion would open a gap.
With SOC, the centro-symmetric phase is already gapped and the distortion instead lowers the gap until a topological phase transition is reached (see Fig. \ref{fig-stability}). Notably, these considerations help to understand the ionic response to an out-of-plane electric displacement field. Such external field not only breaks inversion symmetry in the electronic Hamiltonian through the presence of a linear potential, but it also induces an ionic displacement that breaks the crystal inversion symmetry and lowers the spacegroup from $P\bar{3}m1$ (164) to $P3m1$ (156).
 In Fig. \ref{fig-stability} we report the topological phase diagram, plotting the ionic displacement projected on the  manifold defined by all the possible symmetry-breaking distortions that drive the system from spacegroup 164 to 156 (details in Methods), with respect to an electric displacement field applied orthogonally to the monolayer \cite{lin_pot}. Most of the ionic displacement due to the external field contributes to the crystal inversion-symmetry breaking (from $90\%$ to $100\%$, depending on the field intensity), i.e. the same type of distortion of the $\Gamma$-instability discussed above. The ionic response reduces the critical field $D_c$ needed to close the gap and drive the system to a normal insulating phase (see right panel of Fig. 3 and Supplemental Material \cite{SM_switch}). So, the QSHI phase is robust up to $D_c = 0.36$~V$\cdot{\angstrom}^{-1}$ at the DFT level; the much larger zero-field $G_0W_0$ band gap suggests a larger experimental value, which would also be affected by temperature \cite{monserrat_elphti_16,antonius_tempti_16}. At $D_c$ the gap closes into a Dirac point and for higher fields the system becomes a normal insulator. So, we can posit that the QSHI phase of 2D Jacutingaite is very robust but---thanks to the ionic response---switchable using relatively low out-of-plane electric fields, potentially obtained through a gate voltage of a few volts.

Experiments and technological applications of 2D materials inevitably involve a substrate, potentially affecting certain properties. Although the large band gap and the absence of a band-inversion mechanism already ensure a very robust QSHI phase, we study the effect of encapsulation with hexagonal boron nitride (see Supplemental Material) that is, notably, lattice matched to Jacutingaite. The BN/Pt$_2$HgSe$_3$/BN heterostructure is still a QSHI with a DFT band gap of 0.16 eV, almost identical to the 0.15 eV of an isolated Pt$_2$HgSe$_3$ monolayer. Encapsulation could be useful to protect the monolayer from interactions with oxygen, as in the case of many well known 2D materials \cite{chen_phosphbn_15,kang_heterowafer_17}, although desorption is quite facile (see Supplemental Material \cite{SM_OX}).

In conclusions, our work highlights that a 2D monolayer of the newly discovered mineral Jacutingaite is both a robust and yet switchable QSHI, lattice matched to BN. This finding is even more relevant considering that monolayer Jacutingaite is by far the most outstanding QSHI candidate  that we identified screening more than one thousand novel materials recently proposed as exfoliable \cite{mounet_nanotech_2017}. Jacutingaite is either naturally occurring \cite{cabral_first_obs_08} or easily grown \cite{jacutingaite_exp_12}, providing an optimal platform for studying and exploiting topology-protected physics. 

This  work  was  supported  by  the  MARVEL  NCCR of the Swiss NSF.  We also acknowledge support from the H2020 CoE MaX and from the H2020 EPFL Fellows programme. Simulation time was provided by the Swiss National Supercomputing Centre (CSCS) and PRACE through projects at CSCS and CINECA. The authors would like to thank Dr. Thibault Sohier for many useful discussions and for providing the 2D DFT and DFPT codes.

\begin{thebibliography}{26}
\bibitem{be_def} The binding energy is defined as the ground-state energy difference per area between the 3D bulk and the monolayer, i.e. the formation energy per area of the monolayer with respect to the 3D bulk.
\bibitem{lin_pot} The electric displacement field $D_z$ is the slope of the external linear potential applied with open-boundary conditions along the out-of-plane direction \cite{sohier_2dcode_17}.
\end{thebibliography}
%

\end{document}